\documentstyle[preprint,aps,epsf]{revtex}
\begin{document}
\draft
\preprint{\setlength{\baselineskip}{2.6ex}\hfil
\vbox{\hbox{TRI--PP--97--1}\hbox{hep-ph/9701293}
\hbox{}}}

\title{Testing Time Reversal Invariance in Exclusive Semileptonic $B$ Meson
 Decays}

\author{Guo-Hong Wu$^a$\footnote{\tt{gwu@alph02.triumf.ca}
\vspace{-.15in}}, Ken Kiers$^b$\footnote{\tt{kiers@bnl.gov}
\vspace{-.15in}} and
John N. Ng$^{a}$\footnote{\tt{misery@triumf.ca}}}

\address{$^a$TRIUMF Theory Group\\4004 Wesbrook Mall, Vancouver, B.C.,
V6T 2A3 Canada\\ {\rm and} \\$^b$Department of Physics
\\Brookhaven National Laboratory, Upton, NY 11973-5000, USA}
\maketitle

\begin{abstract}
We demonstrate that polarization measurements in exclusive semileptonic
$B$ decays  are powerful tools for unraveling non-standard model sources
of $T$-violation.
 Measurements of the transverse polarization of
the $\tau$ lepton in the $B \rightarrow D \tau \overline{\nu}$  and
 $B \rightarrow D^* \tau \overline{\nu}$ decays  probe separately
effective scalar and pseudoscalar $CP$-violating four-Fermi interactions,
whereas the $D^*$ polarization in the
$B \rightarrow D^* l \overline{\nu}$ ($l=e$, $\mu$) decay
 is sensitive to an effective right-handed quark current interaction.
Two $T$-odd polarization structures exist involving the
$D^*$ polarization and they can be isolated and studied separately.
An estimate of these $T$-odd effects is also given in the context of
supersymmetric theories.

\end{abstract}

\newpage

 {\it\bf 1. Introduction.}
  Testing the standard model (SM) Cabibbo-Kobayashi-Maskawa (CKM)
\cite{CKM} paradigm of $CP$ violation in hadronic $B$ meson decays
\cite{BCP} is a central theme at future $B$ factories.
The semileptonic decay modes are of particular interest both
because they are relatively clean theoretically and because
they have rather large branching ratios.  They thus provide a means
of accurately measuring the CKM angles, for example~\cite{BCP}.
In this letter we examine some of the exclusive semileptonic decay modes
of the $B$ meson and argue that polarization measurements
in these modes act as sensitive probes of non-standard model
sources of time reversal invariance violation.
We start by classifying the various $T$-violating operators
in terms of effective four-Fermi interactions.  We then calculate
several $T$-odd polarization observables (TOPO's) in terms of these
effective interactions.  As an example, we provide estimates
of these observables arising from  squark family mixings in
supersymmetry (SUSY). $CPT$ symmetry will be assumed throughout this work.

 It is well known that the SM CKM phase has a negligible effect on
$CP$-violating observables in meson
semileptonic decays~\cite{GV}. By way of contrast,
non-SM sources of $CP$ violation could lead to large observable
effects in these decays.
The first experimental attempt to measure the transverse
muon polarization in kaon semileptonic decays \cite{expmu}
has triggered many theoretical discussions of
the muon polarization in both
the $K^+ \rightarrow \pi^0 \mu^+ \nu$ ($K^+_{\mu3}$)
\cite{zhit,cheng,kmu3,CF,wn1} and
$K^+ \rightarrow \mu^+ \nu \gamma $ ($K^+_{\mu2\gamma}$) decays
\cite{marciano,geng,KLO,wn2},
both of which will be measured to a high accuracy at the
on-going KEK E246 experiment \cite{kuno} and at a recently proposed
BNL experiment \cite{adair}.
Similar analyses of the lepton polarization in $D$ and $B$ meson
semileptonic decays have also been discussed in multi-Higgs
models \cite{atwood,garisto,grossman,kunob} and in supersymmetric
theories \cite{wn1}.

  The $B$ mesons have more semileptonic decay channels than
do the kaons, which leads to  the interesting possibility
that different sources of non-SM $CP$ violation
in the $B$ system could be disentangled by performing polarization
measurements in several different decay modes.
Three types of polarization measurements will be discussed in this letter.
The first two involve the lepton transverse polarization
in the $B \rightarrow D l \overline{\nu}$ and
$B \rightarrow D^* l \overline{\nu}$  decays, which
receive contributions from effective
scalar and pseudoscalar interactions, respectively.
Since the lepton polarization involves a chirality flip, we concentrate
on the case $l$$=$$\tau$ in order to take advantage of
the large $\tau$ mass.
The third type of measurement involves the polarization of the vector
meson in the $B \rightarrow D^* l \overline{\nu}$ decay.
The TOPO's  in this case can in general receive
contributions from both an effective pseudoscalar interaction
and from an effective right-handed quark current interaction.
We find the $l=e$, $\mu$ modes to be ideal in searching for a right-handed
current effect -- besides their large branching ratios relative to
the $l=\tau$ mode, they have no pseudoscalar contributions,
due to the small Yukawa coupling of the lepton.
Furthermore, the average $D^*$ polarization
for the $l=e,\mu$ decay can be a factor of three bigger than for
the $l=\tau$ decay.
It is noted that two TOPO's can be constructed
for the $D^*$ polarization, and that they can be separately measured.

We note as an aside that the
electromagnetic final state interactions (FSI)
 in these decays could also lead to
$T$-odd correlation effects \cite{zhit}.
Such effects can be minimized by using charged $B$ decays.
More efficiently,  FSI effects can be
eliminated by taking the difference of the TOPO between
two $CP$-conjugate processes, either for charged or for neutral $B$ mesons.

{\it\bf 2. Form factors.}
  We consider the semileptonic decays $B \rightarrow D l \overline{\nu}$ and
 $B \rightarrow D^* l \overline{\nu}$, where $l=e,\mu,\tau$.
The contributing hadronic matrix elements of the vector and axial-vector
 currents are parameterized by the form factors $f_{\pm}$ and
 $F_{V,A0,A\pm}$ as follows:
\begin{mathletters}
\begin{eqnarray}
\langle D(p^{\prime})| \overline{c} \gamma_{\mu} b |B(p) \rangle
 & = & f_+ \, (p+p^{\prime})_{\mu}
      + f_- \, (p-p^{\prime})_{\mu} \\
\langle D^*(p^{\prime}, \epsilon)| \overline{c} \gamma^{\mu} b |B(p) \rangle
 & = &  i \frac{F_V}{m_B} \epsilon^{\mu\nu\alpha\beta}
         \epsilon^*_{\nu}  (p+ p^{\prime})_{\alpha} q_{\beta} \\
\langle D^*(p^{\prime}, \epsilon)| \overline{c} \gamma_{\mu}
   \gamma_5 b |B(p) \rangle
 & = & - F_{A0} \, m_B \epsilon^*_{\mu}
 -\frac{F_{A+}}{m_B} (p+p^{\prime})_{\mu} \epsilon^* \cdot q
 -\frac{F_{A-}}{m_B} q_{\mu} \epsilon^* \cdot q \, ,
\end{eqnarray}
\label{eq:hme}
\end{mathletters}
where $p$ and $p^{\prime}$ are the four-momenta of the $B$ and $D$
 ($D^*$) respectively, $\epsilon$ is the polarization
vector of the $D^*$ vector meson, $q=p-p^{\prime}$, and the form factors
are functions of $q^2$.  The convention $\epsilon_{0123}=1$ will be used.
The Dirac equation may be applied to the above expressions
in order to obtain the corresponding hadronic matrix
elements for the scalar and pseudoscalar currents.

 The above form factors are relatively real to a good approximation in the SM,
and they are expected to be measured more accurately at the $B$ factories.
At present we must rely on  theoretical models to evaluate them.
For numerical estimates, we adopt the results of heavy quark
effective theory (HQET),  in which
all of the form factors may be expressed in terms of one unknown
Isgur-Wise function $\xi(w)$ \cite{IW},
with $w=\frac{m_B^2+m_{D^{(*)}}^2-q^2}{2m_Bm_{D^{(*)}}}$ and $\xi(1)=1$.

 It is most convenient to parametrize the physics of semileptonic
meson decays by  general effective four-Fermi interactions
of the form,
\begin{eqnarray}
{\cal L} & = & - \frac{G_F}{\sqrt{2}} V_{cb}
        \overline{c} \gamma_{\alpha} (1 - \gamma_5) b
        \overline{l} \gamma^{\alpha} (1- \gamma_5) \nu
       +G_S \overline{c} b \overline{l} (1 - \gamma_5) \nu
   + G_P \overline{c} \gamma_5 b  \overline{l} (1 - \gamma_5) \nu
 \nonumber \\
   & &       + G_R \overline{c} \gamma_{\alpha} (1+\gamma_5)  b
         \overline{l} \gamma^{\alpha} (1- \gamma_5) \nu
   + h.c. , \label{eq:interaction}
\end{eqnarray}
where $G_F$ is the Fermi constant, $V_{cb}$ is the relevant
CKM matrix element, and $G_S$, $G_P$ and $G_R$
denote the strengths of the non-SM interactions due to scalar,
pseudoscalar and right-handed current exchange respectively.
In many models, tensor effects are small compared to the effects
due to the effective interactions of Eq.~(\ref{eq:interaction}).
For simplicity, we neglect tensor interactions in this letter.
 Note that only left-handed neutrinos need to be considered,
as $T$ violation involves the interference between the SM amplitude,
which contains a left-handed neutrino, and the non-SM amplitude.

The effects of new physics on the decay amplitude
are succinctly encapsulated in three dimensionless
 parameters \cite{wn2},
\begin{mathletters}
\begin{eqnarray}
\Delta_S & = &  \frac{\sqrt{2}G_S}{G_FV_{cb}}  \frac{m_B^2}{(m_b-m_c)m_l},
 \\
\Delta_P & = & \frac{\sqrt{2}G_P}{G_FV_{cb}} \frac{m_B^2}{(m_b+m_c)m_l},
 \\
\Delta_R & = & \frac{\sqrt{2}G_R}{G_FV_{cb}},
\end{eqnarray}
\end{mathletters}
where $m_b$, $m_c$, and $m_l$ denote respectively the masses
of the bottom and charm quarks and the mass of the lepton $l$.
Note that $\Delta_S$ and $\Delta_P$ are typically  independent
of the lepton mass because of the leptonic Yukawa coupling
contained in $G_S$ and $G_P$ \cite{rmk}.
These $\Delta$ parameters are in general complex and could
give rise to observable $CP$-violating effects.

{\it\bf 3. $\tau$ polarization.}
   The transverse polarization of the $\tau$ lepton
in the $B\rightarrow D^{(*)} \tau \overline{\nu}$ decays is defined as
\begin{eqnarray}
P^{\bot}_{\tau} & = & \frac{ d\Gamma({\vec{n}}) - d\Gamma(-\vec{n})}
                           { d\Gamma_{total}},
\end{eqnarray}
where ${\vec{n}}$$=$$\frac{\bf{p}_{D^{(*)}} \times \bf{p}_{\tau}}
{|\bf{p}_{D^{(*)}} \times \bf{p}_{\tau}|}$
is a unit vector perpendicular to
the decay plane in the $B$ rest frame,
$d\Gamma(\pm\vec{n})$ is the partial differential
width with the $\tau$ polarization along $\pm\vec{n}$, and
$d\Gamma_{total}$ denotes the partial width summed over the
polarizations.

 In the $B$ rest frame, it is convenient to introduce the variables
$x=2p \cdot p^{\prime}/p^2=2E_{D^{(*)}}/m_B$ and
$y=2p \cdot p_{l}/p^2=2E_{l}/m_B$ as a measure of the
$D^{(*)}$ and lepton energies, as well as two dimensionless quantities
$r_D=m^2_{D^{(*)}}/m_B^2$ and $r_{l}=m^2_{l}/m^2_B$.
The average tau polarization over a region of phase space $S$
may then be defined as
\begin{eqnarray}
\overline{P_{\tau}} & \equiv &
\frac{\int_S dxdy \rho(x,y) P^{\bot}_{\tau}(x,y)}
     {\int_S dxdy \rho(x,y)} \, ,
\label{eq:polav}
\end{eqnarray}
in which $\rho(x,y)$ is proportional to the partial width of the
decay in question,
$\frac{d^2\Gamma(B \rightarrow D^{(*)} \tau \overline{\nu})}{dxdy}  =
 \frac{G_F^2 |V_{cb}|^2 m_B^5}{128 \pi^3}  \rho_{D^{(*)}}(x,y)$.
This average is a measure of the difference between the number
of $\tau$ leptons with their spins pointing above and below the decay plane
divided by the total number of $\tau$ leptons in the same region of
phase space $S$.

The analysis of the $B\rightarrow D \tau \overline{\nu}$ decay proceeds
in complete analogy with that of the $K^+_{\mu3}$ decay \cite{kmu3}.
As the TOPO in this case arises from the interference between
the vector and scalar hadronic matrix elements, its effect will be
directly proportional to the strength of the induced scalar interaction.
In the $B$ rest frame, the transverse polarization of the $\tau$ lepton
is given by
\begin{eqnarray}
P^{\bot(D)}_{\tau}(x,y) & = & - \sigma_D(x,y) Im \Delta_S
\label{eq:poltauD} \\
\sigma_D(x,y) & = & h_D(x)
\frac{\sqrt{r_{\tau}}}{\rho_D(x,y)}
 \sqrt{(x^2-4r_D)(y^2-4r_{\tau}) -4(1-x-y +\frac{1}{2}xy
        +r_D + r_{\tau})^2}
 \nonumber \\
h_D(x) & = & 2 f_+^2 (1-r_D) + 2 f_+ f_- (1-x+r_D) \nonumber \, .
\end{eqnarray}
In HQET, $h_D(x) \rightarrow   (1-r_D)(1+ \frac{x}{2\sqrt{r_D}}) \xi^2$.
Various parameterizations of $\xi(w)$ exist and may
be obtained from the literature~\cite{neupr}.
Notice that whereas $\rho_{D}(x,y)$ has a quadratic dependence on
$\xi$, the polarization function $\sigma_D(x,y)$ is independent of
it.  The average polarization as defined in Eq.~(\ref{eq:polav})
varies slightly for the various forms suggested for $\xi$.
 These comments apply generally to polarization observables,
including those relating to the
$D^*$ polarization to be discussed below.
For numerical estimates we use
the form $\xi(w)=\exp(1-w)$, which is consistent with the
experimental data \cite{opalcleo}.
The average $\tau$ polarization over the whole phase space is then given by
\begin{eqnarray}
\overline{P^{(D)}_{\tau}} & = & - \overline{\sigma_D} Im \Delta_S =
- 0.22 \times Im \Delta_S \, ,
\label{eq:polavtauD}
\end{eqnarray}
which depends only on the non-SM effective scalar interactions.

The transverse polarization of the $\tau$ in the decay
$B \rightarrow D^* \tau \overline{\nu}$
can be similarly calculated,
\begin{eqnarray}
P^{\bot(D^*)}_{\tau} & = & - \sigma_{D^*}(x,y) Im \Delta_P ,
\end{eqnarray}
where $\sigma_{D^*}(x,y)$ is similar in form to $\sigma_D(x,y)$,
with the replacement in Eq.~(\ref{eq:poltauD}) of
$\rho_D(x,y)$ by $\rho_{D^*}(x,y)$
and $h_D(x)$ by a function $h_{D^*}(x)$ which
depends on the three axial form factors.
In HQET, $h_{D^*}(x)$ coincides with $h_D(x)$ to leading order
in the $1/m_{b,c}$ expansion (neglecting QCD corrections).

The $\tau$ polarization in the
$B \rightarrow D^* \tau \overline{\nu}$ decay
is sensitive only to an effective pseudoscalar interaction.
This can be understood by examining the polarization components
 of the $D^*$ involved in the interferences between the vector, axial-vector,
and pseudoscalar hadronic matrix elements.
It can be seen from Eq.~(\ref{eq:hme}) that the vector-axial-vector
interference involves only the two transverse polarizations of the $D^*$,
and that their effects cancel against each other \cite{garisto}.
Therefore the transverse $\tau$ polarization due to this interference
vanishes after summing over the $D^*$ polarizations.
The $\tau$ polarization receives a
non-zero contribution  solely from
the axial-vector-pseudoscalar interference, which involves only the
longitudinal polarization of the $D^*$.
This leads to its dependence on the effective pseudoscalar interaction.

   Averaging over the whole phase space in this case gives
\begin{eqnarray}
\overline{P^{(D^*)}_{\tau}} & = & - \overline{\sigma_{D^*}} Im \Delta_P
= - 0.067 \times Im \Delta_P \, .
\label{eq:polavtauD*}
\end{eqnarray}
Comparing Eqs.~(\ref{eq:polavtauD}) and (\ref{eq:polavtauD*}),
we see that the coefficient of the latter is smaller by about a factor
of three.  This may be understood roughly by looking at the
definition of the average polarization given in Eq.~(\ref{eq:polav})
and noting that, for the $B \to D^* \tau \overline{\nu}$ decay,
effectively only one of the three polarization states of the $D^*$
contributes to the numerator whereas all three contribute to the denominator.

{\it\bf 4. $D^*$ polarization.}
  We now consider the $D^*$ polarization in the
$B \rightarrow D^* l \overline{\nu}$ decay.
The polarization of the $D^*$ can be measured by studying the angular
distribution of its decay products in the decays $D^* \to D \pi$ and
$D^* \to D \gamma$.
$CP$-violating effects can therefore also be studied by
considering $T$-odd momentum correlations in a specific
four-body final state of the $B$ decay \cite{GV,KSW}.
To be completely general, we present our calculation in terms
of the $D^*$ polarization vector.  The resulting expressions are then
generically applicable to pseudoscalar decays of the type
$P$$\to$$Vl\overline{\nu}$, in which the vector meson $V$ may decay
differently than the $D^*$.

Working in the $B$ rest frame, we
denote the three-momenta of the $D^*$ and $l$
by ${\bf p}_{D^*}$ and ${\bf p}_{l}$, and define three orthogonal
vectors
${\vec{n}}_1  \equiv  \frac{({\bf p}_{D^*} \times {\bf p}_{l})
    \times {\bf p}_{D^*}}
     {|({\bf p}_{D^*} \times {\bf p}_{l})
    \times {\bf p}_{D^*}|}$,
${\vec{n}}_2  \equiv  \frac{{\bf p}_{D^*} \times {\bf p}_{l}}
                     {|{\bf p}_{D^*} \times {\bf p}_{l}|}$,
and
${\vec{n}}_3= \frac{{\bf p}_{D^*}}{|{\bf p}_{D^*}|} \frac{m_{D^*}}{E_{D^*}}$.
The vector ${\vec{n}}_3$ has been chosen such that
the constraint $\epsilon^2=-1$ becomes symmetric in the ${\vec{n}}$'s;
i.e.
  $(\vec{\epsilon}\cdot {\vec{n}}_1)^2 +
  (\vec{\epsilon}\cdot {\vec{n}}_2)^2 +
   (\vec{\epsilon}\cdot {\vec{n}}_3)^2=1$.
   The polarization vectors of the $D^*$ can be taken to be real,
corresponding to plane polarizations.
In this linear basis,
the TOPO's have a clear physical interpretation
and can be easily constructed.
Note that the $D^*$ polarization projection transverse to the decay plane,
$\vec{\epsilon}\cdot {\vec{n}}_2$, is $T$-odd, while those
inside  the decay plane,
$\vec{\epsilon}\cdot {\vec{n}}_1$ and $\vec{\epsilon}\cdot \vec{n}_3$,
are $T$-even.
 TOPO's arising from interference between
different amplitudes should therefore involve the product
$(\vec{\epsilon}\cdot \vec{n}_2)(\vec{\epsilon}\cdot {\vec{n}}_1)$
or $(\vec{\epsilon}\cdot {\vec{n}}_2)
(\vec{\epsilon}\cdot {\vec{n}}_3)$.

  A measure of the $T$-odd correlation involving the $D^*$ polarization
can be defined as
\begin{eqnarray}
P_{D^*}(x,y) & \equiv &
 \frac{d\Gamma - d\Gamma^{\prime}}{d\Gamma_{total}} =
  \frac{2d\Gamma_{T-odd}}{d\Gamma_{total}} ,
\label{eq:polD*}
\end{eqnarray}
where $d\Gamma^{\prime}$ is obtained by performing a $T$ transformation
on  $d\Gamma$,
 $d\Gamma_{T-odd}$ is the $T$-odd piece in the partial width,
and $d\Gamma_{total}$ is the partial width summed over
 $D^*$ polarizations.

  It can be shown \cite{wkn} that both
the $G_R$ and $G_P$ interactions of Eq.~(\ref{eq:interaction})
contribute to $P_{D^*}$ of Eq.~(\ref{eq:polD*}),
and that the $G_P$ effect is Yukawa-suppressed by $r_l=m_l^2/m_B^2$.
Thus by using the $e$ or $\mu$ mode of the $B$ decay,
the $G_R$ effect can be isolated  and measured.
To a good approximation, we may neglect the masses of the electron and muon
by  setting $r_l=m_l^2/m_B^2=0$.
The ${D^*}$ polarization for the $B\rightarrow D^* l\overline{\nu}$
($l=e,\mu$) decay is then simply given by
\begin{eqnarray}
P_{D^*}(x,y) & = &
 \left(\sigma_1(x,y) (\vec{\epsilon}\cdot {\vec{n}}_1)
  + \sigma_2(x,y) (\vec{\epsilon}\cdot {\vec{n}}_3) \right)
  (\vec{\epsilon}\cdot {\vec{n}}_2) Im \Delta_R \, ,
\label{eq:D*pol}
\end{eqnarray}
with the two polarization functions defined as
\begin{mathletters}
\begin{eqnarray}
\sigma_1(x,y) & = & -
\frac{
     8(x^2-4r_D)y^2-32(1-x-y +\frac{1}{2}xy
        +r_D)^2}{\rho_{D^*}(x,y)\sqrt{x^2-4r_D}}
 F_{A0} F_V \\
\sigma_2(x,y) & = &
 \frac{8(x+2y-2)\sqrt{(x^2-4r_D)y^2 -4(1-x-y +\frac{1}{2}xy +r_D)^2}}
      {\rho_{D^*}(x,y)\sqrt{4r_D(x^2-4r_D)}} \nonumber \\ &&
 \times F_V \left(F_{A+} (x^2-4r_D) + F_{A0} (x-2r_D) \right)  \, .
\end{eqnarray}
\end{mathletters}

  The two terms in the $D^*$ polarization of Eq.~(\ref{eq:D*pol})
have quite different polarization structures. Note that the
term proportional to $\sigma_1$ involves transverse components of the
polarization vector only,
while the term proportional to $\sigma_2$ contains both transverse and
longitudinal polarization components and
would be absent for on-shell massless vector bosons
such as the photon.
Using symmetry properties of $\sigma_1$ and $\sigma_2$, two TOPO's can be
constructed which correspond separately to the two polarization
structures of  Eq.~(\ref{eq:D*pol}).

In order to isolate the $\sigma_1$ term of Eq.~(\ref{eq:D*pol}),
we observe that $\rho_{D^*}(x,y)\sigma_2(x,y)$
is antisymmetric under the exchange of lepton and anti-neutrino
energies and that the allowed phase space is symmetric under the same
exchange. Thus, performing the average in Eq.~(\ref{eq:polav})
over the whole phase space (or any
region $S$ of the phase space which is symmetric in the lepton and
neutrino energies) eliminates the $\sigma_2$ term and
leaves only the first polarization structure:
\begin{eqnarray}
 \left. \overline{P^{(1)}_{D^*}}\right|_{all} &\simeq & 0.51 \times
(\vec{\epsilon}\cdot {\vec{n}}_1)(\vec{\epsilon}\cdot \vec{n}_2)
Im \Delta_R .
\label{eq:TOPO1}
\end{eqnarray}
This simple form is a consequence of the near masslessness of the lepton
(i.e. $r_l \simeq 0$) and the symmetry of the integration region.
It is  valid independent of the functional form of the form
factors, which depend only on $x$.

The second polarization structure may be separated out by
making use of the reflection symmetry
of $\rho_{D^*}(x,y)\sigma_1(x,y)$ under $y \to 2-x-y$ and $x \to x$.
This symmetry amounts to reflecting the lepton energy with respect to its
mid-point value $y_{mid}=(y_{min} + y_{max})/2=1-x/2$ for a given $x$.
We can thus define the following asymmetric average over the whole
phase space to eliminate the $\sigma_1$ term,
\begin{eqnarray}
  \left. \overline{P^{(2)}_{D^*}}\right|_{all}
& \equiv & \frac{ \int dx \left( \int_{y_{min}}^{y_{mid}} dy
                               -\int_{y_{mid}}^{y_{max}} dy \right)
     \rho_{D^*}(x,y) P_{D^*}(x,y) }
    {\int dx dy \rho_{D^*}(x,y)}  \nonumber \\
 &\simeq & 0.39 \times
(\vec{\epsilon}\cdot {\vec{n}}_2)(\vec{\epsilon}\cdot \vec{n}_3)
Im \Delta_R .
\label{eq:TOPO2}
\end{eqnarray}
Unlike the TOPO of Eq.~(\ref{eq:TOPO1}),
 this method of isolating the transverse-longitudinal
polarization structure works independent of the lepton mass.
In fact, this procedure also eliminates the
pseudoscalar contribution which is generally present for non-zero
lepton masses.
We find, however, that the average $D^*$ polarization for the $\tau$ mode
using this prescription
is about a factor of three smaller than that for the electron or muon mode.

 The two observables  of Eqs.~(\ref{eq:TOPO1}) and (\ref{eq:TOPO2})
can be separately related to  two $T$-odd momentum correlation observables
in the four-body final state $B \to D^*(D\pi) l \overline{\nu}$ \cite{wkn}.
Complementary measurements of both observables may then be used to
provide a consistency check regarding
the possible existence of a right-handed current effect.

{\it\bf 5. SUSY effects.}
As an example, we now estimate the size of the TOPO's from
squark family mixings in SUSY.
In the supersymmetric standard model with $R$-parity conservation,
mass matrices of the quarks and squarks are generally expected
to be diagonalized by different unitary transformations.
The relative rotations in generation space between the
$\tilde{u}_L$, $\tilde{u}_R$, $\tilde{d}_L$, and $\tilde{d}_R$
squarks and their corresponding quark partners are denoted by
$V^{U_L}$, $V^{U_R}$,$V^{D_L}$, and $V^{D_R}$ respectively.
These mixing matrices appear in the quark-squark-gluino couplings,
and the products of different $V^U$'s ($V^D$'s)
are constrained by flavor-changing-neutral-current (FCNC) processes
in the up (down) quark sector \cite{fcnc}.
Charged current processes, on the other hand, involve products
of $V^U$ and $V^D$ which may be of order unity
without violating the experimental FCNC bound \cite{wn1,wn2}.
We consider only top and bottom squark loop diagrams in our calculation,
as they tend to dominate in the presence of  large squark-generational
mixings.
The relevant mixing matrix elements involved in $B$ meson semileptonic
decays are $V^{U}_{32}$ and $V^{D}_{33}$.
For simplicity, we consider only phases in the mixing matrices
and ignore the phases in other SUSY parameters. This is sufficient
for an estimate of the maximal $T$-odd polarization effects.

  The leading contribution to $\Delta_S$ and $\Delta_P$ comes
from the charged Higgs exchange diagram at one loop \cite{wn2}.
The $m_t$-enhanced effect is given by
\begin{eqnarray}
\Delta_S & = & - \frac{\alpha_s}{3\pi} I_H \tan \beta
\frac{m_B}{(m_b-m_c)} \frac{m_B m_t}{m_H^2} \times
\frac{\mu + A_t\cot\beta}{m_{\tilde{g}}} \times
\frac{[V^{H}_{33} V^{D_L}_{33} {V^{U_R}_{32}}^*]}{V_{cb}} , \\
\Delta_P & = &  \frac{\alpha_s}{3\pi} I_H \tan \beta
\frac{m_B}{(m_b+m_c)} \frac{m_B m_t}{m_H^2} \times
\frac{\mu + A_t\cot\beta}{m_{\tilde{g}}} \times
\frac{[V^{H}_{33} V^{D_L}_{33} {V^{U_R}_{32}}^*]}{V_{cb}} ,
\label{eq:deltaSP}
\end{eqnarray}
where $\alpha_s\simeq 0.1$ is the QCD coupling
evaluated at the mass scale of the sparticles in the loop,
$A_t$ is the soft SUSY breaking $A$ term for the top
 squark, $\mu$ denotes the two Higgs superfields mixing
parameter, $\tan \beta$ is the ratio of the two Higgs VEVs,
$m_{\tilde{g}}$ is the mass of the  gluino and
$V^{H}_{ij}$ is the mixing matrix in the charged-Higgs-squark
coupling $H^+ {\tilde{u_i}_R}^* \tilde{d_j}_L$.
$I_{H}$ is  an integral function of order one, defined previously
in Refs.~\cite{wn1,wn2}.

To estimate  the maximal allowed SUSY effect,
we assume $|V^{D_L}_{33}|=|V^{H}_{33}| \sim 1$,
and take $m_{H}=100\; \mbox{GeV}$ and
$\tan \beta=50$ to saturate the bound
$\tan \beta /m_H < 0.5$ GeV$^{-1}$
from the $b\rightarrow c \tau \overline{\nu}$
decay \cite{GHN}.  With maximal squark mixings,
$|V^{U_R}_{32}|=\frac{1}{\sqrt{2}}$ \cite{worah}.
Setting $|\mu|=A_t=m_{\tilde{g}}$ and taking $m_b=4.5\;\mbox{GeV}$,
$m_c=1.5\;\mbox{GeV}$, and $I_{H}=1$, we find
$|\Delta_S|  \le  1.6$ and
$|\Delta_P|  \le 0.78$.
 The $\tau$ polarization  in the
$B \rightarrow D \tau \overline{\nu}$ decay is then
$\left|\overline{P^{(D)}_{\tau}}\right| \le 0.35$,
and for $B \rightarrow D^* \tau \overline{\nu}$ decay
$\left|\overline{P^{(D^*)}_{\tau}}\right| \le 0.05$.
 Both limits scale as $ \left( \case{100\text{ GeV}}{m_H} \right)^2
                \frac{\tan \beta}{50}
   \frac{Im[V^{H}_{33} V^{D_L}_{33} {V^{U_R}_{32}}^*]}{\sqrt{2}/2}$.

   An effective $G_R$ interaction can be induced
 at one loop by the $W$-boson exchange with left-right mass insertions
 in  both the top and bottom squark propagators.
In this insertion approximation,
  the $\Delta_R$ parameter is computed to be \cite{wn2}
\begin{equation}
\Delta_R =
 - \frac{\alpha_s}{36 \pi} I_0
         \frac{m_tm_b(A_t-\mu\cot\beta)(A_b-\mu\tan\beta)}
              {m_{\tilde{g}}^4}
 \frac{V^{SKM}_{33}{V^{U_R}_{32}}^*V^{D_R}_{33}}{V_{cb}} ,
\end{equation}
where $A_b$ is the soft SUSY breaking $A$ term for the bottom squark,
and $V^{SKM}_{ij}$ is the super CKM matrix associated with the
$W$-squark coupling $W^+ {\tilde{u_i}_L}^* \tilde{d_j}_L$.
The integral function $I_0=1$ for
$m_{\tilde{t}_{L,R}}= m_{\tilde{b}_{L,R}}=m_{\tilde{g}}$,
but can be of order 10 for reasonable squark-gluino mass splittings
\cite{wn2}.
To estimate the maximal size of $\Delta_R$, we take
$I_0=5$, $\tan \beta=50$, $A_t=A_b=|\mu|=m_{\tilde{g}}=200 \;
\mbox{GeV}$, $|V^{D_R}_{33}|=|V^{SKM}_{33}|=1$,
and $|V^{U_R}_{32}|=\frac{1}{\sqrt{2}}$ for maximal
squark family mixings. For $|V_{cb}|=0.04$, we have
$|\Delta_R|  \le  0.08$.
Therefore the average polarizations of the $D^*$ over all phase space
are simply given by
 $\left| \overline{P^{(1)}_{D^*}}\right|_{all} < 0.02$
and $\left| \overline{P^{(2)}_{D^*}}\right|_{all} < 0.016$,
where the optimal case occurs when
$(\vec{\epsilon}\cdot {\vec{n}}_i)(\vec{\epsilon}\cdot \vec{n}_j)=1/2$
 ($i\ne j$).
These limits scale as
$\frac{I_0}{5}
\left( \frac{200\text{ GeV}}{M_{SUSY}} \right)^2
 \left( \frac{\tan \beta}{50} \right)
 \left( \frac{Im[V^{SKM}_{33}{V^{U_R}_{32}}^*V^{D_R}_{33}]}{\sqrt{2}/2}
\right)$, where $M_{SUSY}$ is the SUSY breaking scale.

{\it\bf 6. Conclusion.}
  The $T$-odd polarization observables
constructed and estimated in this letter can be separately measured
to identify the sources of non-SM $T$ violation in semileptonic
$B$ decays, or to put bounds on them.
A model-independent comparison can be made among the various TOPO's
by combining the averages given
in Eqs.~(\ref{eq:polavtauD},\ref{eq:polavtauD*},\ref{eq:TOPO1},\ref{eq:TOPO2})
with the branching ratios for the various decay modes.
 It is particularly interesting to note that the decay
$B \to D^*l\overline{\nu}$ ($l=e,\mu$) has a branching ratio which is about
10 times that of the $B \to D \tau \overline{\nu}$ mode, so that
by including both the electron and muon modes in the $D^*$ polarization
measurement, one can expect an event rate
which is about 20 times higher than
that for the $\tau$ polarization measurement in $B \to D \tau \overline{\nu}$
decay.  It thus appears that polarization measurements in $B$ decays
may generically have a better sensitivity to a
$CP$-violating right-handed current effect ($\Delta_R$)
than to a scalar ($\Delta_S$) or pseudoscalar ($\Delta_P$) interaction.

We have also estimated
the maximal size of the various polarization
observables from squark family mixings in SUSY.
Although these effects arise at the one-loop level,
they are not necessarily small if
a large mixing between the right-handed charm and top squarks
($V^{U_R}_{32}$) exists. The induced scalar effect
seems to have a better chance for
detection at the planned $B$ factories than do the induced
pseudoscalar or right-handed current effects.

\acknowledgments
 We would like to thank S. Dawson,  F. Goldhaber, Y. Kuno, T. Numao,
and A. Soni for helpful discussions.
 This work is partially supported by the Natural Sciences and
Engineering Research Council of Canada.  K.K.
is also grateful to Brookhaven National Laboratory for support
provided under contract number DE-AC02-76CH00016 with the U.S.
Department of Energy.

\addtolength{\baselineskip}{-.3\baselineskip}


\begin{thebibliography}{10}

\bibitem{CKM}
N. Cabibbo, Phys. Rev. Lett. 10 (1963) 531;\\
M.~Kobayashi and T.~Maskawa, Prog. Theor. Phys. 49 (1973) 652.

\bibitem{BCP}
For a review, see for example, {\it B Decays}, ed. S. Stone, 1994.

\bibitem{GV}
E. Golowich and G. Valencia, Phys. Rev. D 40 (1989) 112.

\bibitem{expmu}
M.K. Campbell {\it et al.} Phys. Rev. Lett. 47 (1981) 1032;\\
S.R. Blatt {\it et al.}, Phys. Rev. D 27 (1983) 1056.

\bibitem{zhit}
A.R. Zhitnitskii, Yad. Fiz. 31 (1980) 1024
[Sov. J. Nucl. Phys., 31 (1980) 529].

\bibitem{cheng}
M. Leurer, Phys. Rev. Lett. 62 (1989) 1967; \\
P. Castoldi, J.M. Frere and G. Kane, Phys. Rev. D 39 (1989) 263;\\
H.Y. Cheng, Phys. Rev. D 28 (1983) 150.

\bibitem{kmu3}
H.Y. Cheng, Phys. Rev. D 26 (1982) 143; \\
R. Garisto and G. Kane, Phys. Rev. D 44 (1991) 2038; \\
G. Belanger and C.Q. Geng, Phys. Rev. D 44 (1991) 2789.

\bibitem{CF}
E. Christova and M. Fabbrichesi, Phys. Lett. B 315 (1993) 113; \\
M. Fabbrichesi and F. Vissani, hep-ph/9611237.

\bibitem{wn1}
G.-H. Wu  and J.N. Ng, hep-ph/9609314, to appear in Phys. Lett. B.

\bibitem{marciano}
W.J. Marciano,
in the Proceedings of ``Workshop on Future Directions
in Particle and Nuclear Physics at Multi-GeV Hadron Beam
Facilities,'' BNL, 1993.

\bibitem{geng}
C.Q. Geng and S.K. Lee, Phys. Rev. D 51 (1995) 99.

\bibitem{KLO}
M. Kobayashi, T-T. Lin, and Y. Okada, Prog. Theor. Phys.
95 (1996) 361.

\bibitem{wn2}
G.-H. Wu  and J.N. Ng, Phys. Rev. D 55 (1997) 2806, hep-ph/9610533.

\bibitem{kuno}
Y. Kuno, Nucl. Phys. B (Proc. Suppl.) 37A (1994) 87.

\bibitem{adair}
M.V. Diwan  {\it et. al.}, AGS Proposal - Search for $T$ Violating
 Muon Polarization in $K^+ \rightarrow \mu^+ \pi^0 \nu$ Decay,
 July, 1996.

\bibitem{atwood}
D. Atwood, G. Eilam and A. Soni, Phys. Rev. Lett. 71 (1993) 492.

\bibitem{garisto}
R. Garisto, Phys. Rev. D 51 (1995) 1107.

\bibitem{grossman}
Y. Grossman and Z. Ligeti, Phys. Lett. B 347 (1995) 399.

\bibitem{kunob}
Y. Kuno, Chinese J. Phys. 32 (1994) 1015.

\bibitem{IW}
N. Isgur and M. Wise, Phys. Lett. B 232 (1989) 113;
237 (1990) 527.

\bibitem{rmk}
It is possible that $G_S$ and $G_P$ are not directly
proportional to the lepton mass as in SUSY $R$-parity breaking models
and leptoquark models. 
However, the size of $G_S$ and $G_P$ involving the light leptons 
is often severely constrained by precision measurements in low 
energy processes \cite{wkn}.

\bibitem{neupr}
M. Neubert, Phys. Rep. 245 (1994) 259;\\
J.D. Richman and P.R. Burchat, Rev. Mod. Phys. 67 (1995) 893.

\bibitem{opalcleo}
CLEO Collaboration, J.E. Duboscq {\it et al.}, Phys. Rev. Lett.
76 (1996) 3898;\\
OPAL Collaboration, K. Ackerstaff {\it et al.}, CERN-PPE/96-162.

\bibitem{KSW}
J.G. K\"{o}rner, K. Schilcher and Y.L. Wu, Phys. Lett. B 242 (1990) 119.

\bibitem{wkn}
G.-H. Wu, K. Kiers and J.N. Ng, in preparation.

\bibitem{fcnc}
F. Gabbiani, E. Gabrielli, A. Masiero, and L. Silvestrini, hep-ph/9604387,
Nucl. Phys. B477 (1996) 321.

\bibitem{GHN}
Y. Grossman, H. Haber and Y. Nir, Phys. Lett. B 357 (1995) 630;\\
Y. Grossman and Z. Ligeti, Phys. Lett. B 332 (1994) 373.

\bibitem{worah}
The phenomenological implications of
 $|V^{U_R}_{32}|=1/\sqrt{2}$ have been discussed in the
context of $CP$ violation in the neutral meson mixings
by M.P.~Worah, Phys. Rev. D 54 (1996) 2198.

\end{thebibliography}
\end{document}